\documentstyle[11pt,moriond,epsfig]{article}

\def\Journal#1#2#3#4{{#1} {\bf #2}, #3 (#4)}

\def\PLB{{\em Phys. Lett.}  B}
\def\PRL{\em Phys. Rev. Lett.}
\def\PRC{{\em Phys. Rev.} C}
\def\PRD{{\em Phys. Rev.} D}

\def\be{\begin{equation}}
\def\ee{\end{equation}}
\def\bea{\begin{eqnarray}}
\def\eea{\end{eqnarray}}

\begin{document}
\vspace*{4cm}
\title{The experimental future of Neutrino Oscillations}

\author{Mario Campanelli}

\address{Institut f\"ur Teilchenphysik ETH Z\"urich Switzerland}

\maketitle\abstracts{
After the recent experimental results on neutrino oscillations, some shape
starts to emerge from the puzzle. However, the situation is still far from
being clarified. First of all, accommodating all experimental
results in a single and simple framework is not possible, and the possibility
of sterile neutrinos entering the oscillation process has not been ruled out.
Moreover, new questions arise that the presently-available data, nor those
that will be available in a near future, will be able to answer.
In this paper some of these problems will be discussed, as well as the 
experimental guidelines for their clarification.
} 

\section{Introduction}
At present, there are three classes of experiments where effects that could
be interpreted as indications of neutrino oscillations have been seen. They 
are,
in decreasing order of robustness of the result:\begin{itemize}
\item {\bf Atmospheric neutrinos} Several different experiments \cite{atm}
are consistent
in indicating an angle-dependent disappearance of muon neutrinos, with
maximal mixing and a mass difference squared of the order of $\Delta m^2
\approx 3\times 10^{-3} eV^2$. The angular dependence of the deficit is a 
clear smoking gun for oscillations, since it is incompatible with other
effects (i.e. neutrino decay).
\item {\bf Solar neutrinos} A large deficit in the number of electron 
neutrinos observed with respect to the standard solar model has been 
found by several experiments in different energy ranges\cite{solar}. It can
be interpreted as oscillations involving electron neutrinos with a mass
difference squared around $10^{-5}, 10^{-7}$ or $10^{-10} eV^2$ (with
the first value preferred by the latest results)
Independent 
measurements on the solar activity (like heliosismology lines) seem to 
validate the robustness of the solar model, but no unambiguous indication
for oscillation has been found. For space reasons, future experiments for
solar neutrino detections will not be discussed here.
\item {\bf Accelerator neutrinos (LSND)} An excess of electron events in a $\nu_\mu$ beam
has been observed by the LSND \cite{lsnd} collaboration for over 5 years.
This result was not confirmed by the KARMEN II \cite{karmen2} 
experiment, but a small region of the parameter space is still allowed by the
combination of the two experiments, for $\nu_\mu\to\nu_e$ transitions
with a mass difference squared around $0.1-1.0 eV^2$.
\end{itemize}
The data from atmospheric neutrinos largely favor $\nu_\mu\to\nu_\tau$ 
transitions; if however the $\nu_e$ disappearance in solar neutrinos is 
interpreted as result of
$\nu_e\to\nu_\mu$ transitions, the allowed parameter space
would be very far from the indications of the LSND 
experiment. In any case, no matter how data are 
interpreted, the existence of three independent mass differences is not 
compatible with oscillation involving only three neutrino families, since
in this case only two independent mass differences would be involved.
If all experimental results have to be explained in terms of neutrino
oscillations, it is mandatory to assume the existence of a fourth neutrino,
that would be sterile, i.e. would have a very small coupling with the Z
boson, otherwise its existence would have emerged from LEP data.
\par
\section{Testing the LSND result}
As we have seen, out of the three present indications for oscillations, the
LSND claim still awaits independent confirmation. And since its 
interpretation as neutrino oscillations requires the introduction of the
sterile states, it is obvious that confirming or disproving LSND is one
of the crucial issues of future experimental neutrino physics. This will
be the main goal of the MiniBOONE experiment \cite{mboone}, expected to start
data taking in December 2001. Neutrinos will be produced by protons from 
the Fermilab booster ($<E_p>$=8 GeV), and will have a broad spectrum around
$E_\nu$=1 GeV. The detector will be a large sphere of 807 (445 fiducial) tons 
of scintillating material (mineral oil), read by 1280 8-inch phototubes. 
Like LSND, the aim is to search
for electron appearance in a beam primarily composed of $\nu_\mu$; however
there are several differences among the two experiments: MiniBOONE will
use a beam of approximately 30 times more energy and it will be located at
about 20 times the distance of LSND; the Cerenkov light is
4 times larger than the scintillation light, and particle identification 
will be based on ring shape rather than relying on delayed neutron capture.
The philosophy of this experiment is to aim at large electron signals in
case LSND is correct; however, also the total number of events is large
($\approx$ 600000 CC events/year) and so are the backgrounds (see table
\ref{tab:bgmb} for details). After one year of running, the experiment will be
able to either confirm or disprove the LSND result, hopefully solving by
the beginning of 2003 one of the most important open problems in neutrino
physics.
\begin{table}[tbh]
\caption{Main background sources for the MiniBOONE experiment\label{tab:bgmb}}
\vspace{0.4cm}
\begin{center}
\begin{tabular}{|c|c|}
\hline
Source&N. events/year\\ \hline
Misidentified pions& 600\\
Misidentified muons& 600\\
Intrinsic $\nu_e$ in beam& 1800\\
\hline
\end{tabular}
\end{center}
\end{table}

\section{The atmospheric region}
The parameter space region with large mixing angle and mass difference squared
of the order of $3\times 10^{-3} eV^2$ is usually referred to as the 
atmospheric region, since the most natural explanation to the atmospheric 
neutrino anomaly is an oscillation governed by these parameters. In order
to improve the already good present data, two approaches can be followed:
\begin{itemize}
\item a different ``beam'': the uncertainty on atmospheric neutrino production
can be reduced using an artificial beam of neutrinos produced by an 
accelerator, with a baseline of several hundreds of kilometers
\item different detectors: it is possible to aim at lower thresholds, better
L/E resolution, more mass
\end{itemize}
The first long-baseline neutrino beam, the Japanese project K2K, has started
data taking in 1999. For more details about this program and its results,
refer to \cite{k2k}. The other two beams are the American project NuMi from
Fermilab to the Soudan site in Minnesota and the European CNGS from CERN to 
Gran Sasso. Their design is driven by to different philosophies, due
to different physics goals and detector designs. The American experiment aims 
at a precision measurement of the oscillation parameters; a near detector is
planned, and the beam has a tunable energy to have more events where the 
maximum of the oscillation should take place. On the contrary, the European 
experiment will be entirely devoted to $\tau$ search to confirm the 
$\nu\mu\to\nu_\tau$ nature of the atmospheric oscillations, and the beam 
profile will be tuned to produce the largest number of $\tau$ neutrinos in the
detector. 
\subsection{MINOS}
The MINOS detector will be a 5 kton coarse magnetized iron-scintillator 
apparatus, consisting of 486 layers of 2.54 cm thick iron slabs, interleaved
by 1 cm thick scintillator strips read out by wavelength shifter fibers.
Overall, there will be 25800 $m^2$ of active detector planes. The main
measurement will be the precise determination of $\Delta m^2_{23}$ and
$\theta_{23}$ from $\nu_\mu$ disappearance. The maximum of oscillations
occurs for $1.27 \Delta m^2 L/E_\nu = \pi/2$,
corresponding to about 2 GeV for the baseline chosen and the central value
of the atmospheric neutrinos, so the low-energy option for the beam has been
chosen. 
\begin{figure}[tb]
 \begin{minipage}{.45\linewidth}
  \begin{center}
   \epsfig{file=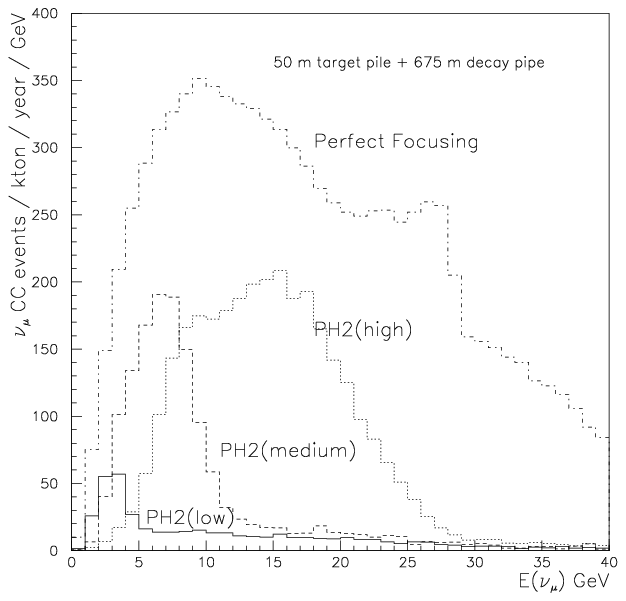,width=8cm,clip}
  \end{center}
\caption{The three possible profiles of the NuMI beam. Given the present value 
of $\Delta m^2_{23}$, the low energy beam will be used.}
\label{fig:minosbeam}
\end{minipage}
 \begin{minipage}{.45\linewidth}
  \begin{center}
   \epsfig{file=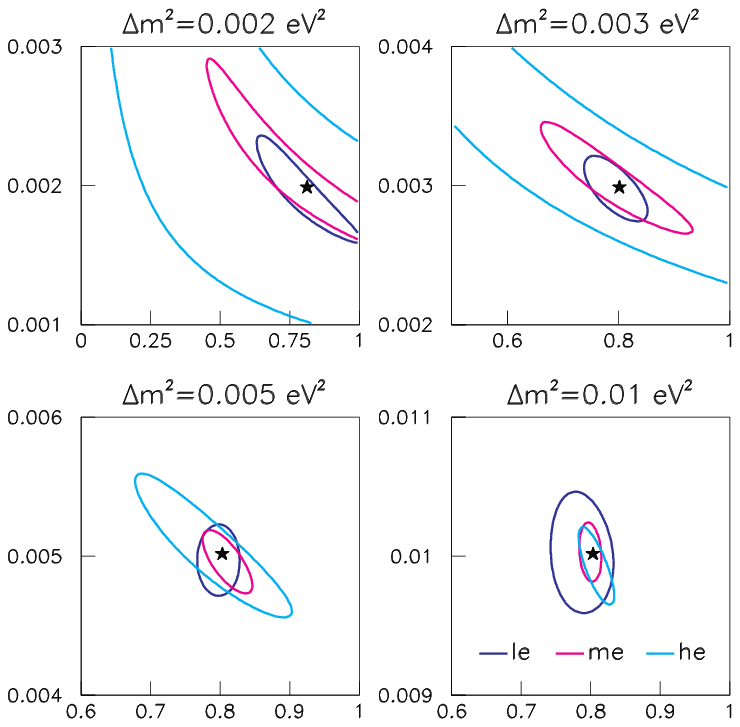,width=7cm,clip}
  \end{center}
\caption{Precision on the determination of $\Delta m^2_{23}$ and $\sin^2 
2\theta_{23}$ for various values of the first parameter, setting the second at
0.8}
\label{fig:minosprec}
\end{minipage}
\end{figure}
The beam profile, together with the two further possible options,  are shown 
in figure \ref{fig:minosbeam}.
The precision achievable in the oscillation parameters $\Delta m^2_{23}$ and
$\sin^2 2\theta_{23}$ is in \ref{fig:minosprec}.\par
The appearance search program from MINOS suffers from the poor detector
granularity; $\nu_\tau\leftrightarrow\nu_s$ discrimination is possible on a 
statistical basis from the ratio of charged current-like and neutral 
current-like events. Also the sensitivity for $\nu_\mu\to\nu_e$ searches is
limited by the detector electron identification capabilities, and for this
measurement MINOS will be used in conjunction with the present SoudanII
apparatus (that in this contest will be called THESEUS \cite{theseus}).
\subsection{OPERA}
\begin{figure}
\begin{center}
\epsfig{file=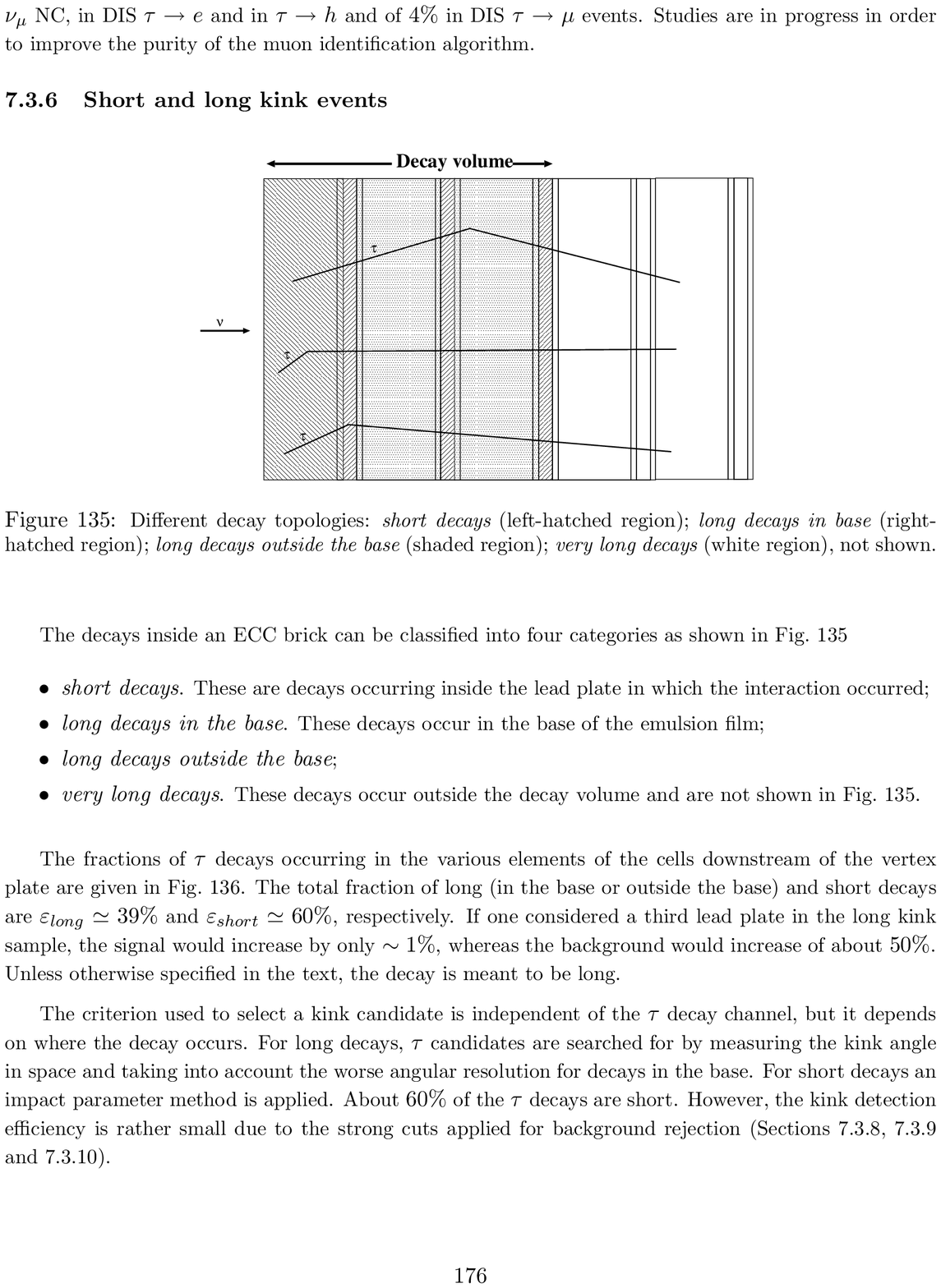,width=8cm,bb=125 501 440 680,clip}
\end{center}
\caption{Possible $\tau$ decays in the OPERA lead-emulsion sandwich. From top:
long decay, long decay in emulsion base, short decay.}
\label{fig:operacell}
\end{figure}

OPERA will be a dedicated experiment to look for $\tau$ appearance at the
CNGS beam. Its design philosophy is to have an almost background-free
experiment, such that even with few candidates it would be possible to claim
discovery of $\nu_\mu\to\nu_\tau$ oscillations. To achieve this goal, the 
$\tau$ will be identified via the kink produced by its decay, exploiting the
very high space resolution provided by nuclear emulsions. Due to their price, 
however, it is not possible to envisage a detector entirely made of emulsions;
it will be made of a passive material (lead) used as a neutrino
target, while the emulsion sheets will be used for tracking (see figure
\ref{fig:operacell}). $\tau$ production will be searched for in three decay
channels: $\tau\to e$, $\tau\to\mu$ and $\tau\to h$. The $\tau$ produced 
from neutrino interactions will either decay in the same lead block 
(short decay, only used in the $\tau\to e$ channel) or in the neighboring one
(long decay). In the first case, they are identified from tracks having 
large impact parameter with respect to the reconstructed vertex; in the
second case, directly from the kink of the track. The total efficiency
($\times$ BR) is expected to be 8.7\%. Main background sources are:
\begin{itemize}
\item cosmic rays and radioactivity from the rocks
\item hadronic decays and re-interactions
\item muon scattering
\item charm decays
\end{itemize}
for an expected total of 0.57 events for 5 years of data taking. In the same 
period, the number of $\tau$ events expected is 4.1, 18.3 and 44.1 for 
$\Delta m^2_{23}= 1.5, 3.2, 5.0 \times 10^{-3} eV^2$, respectively\cite{opera}.
\subsection{ICANOE}
Another approach to the $\tau$ search at the CNGS is proposed by the ICANOE
experiment. The detector will be composed of four modules of large (1245 ton
fiducial mass each) liquid Argon Time 
Projection Chambers, completed with an external muon spectrometer. Its very 
good imaging, particle identification and calorimetric capabilities offer a
wide variety of physics possibilities. One of the main motivations for
developing such a technology is the search for nucleon decay, for which a
background-free search is possible, in particular for the SUSY-preferred
channels (like $p\to K^+ \bar{\nu}$). The detector is a very good next 
generation atmospheric neutrino 
experiment: it allows detection of all neutrino flavors, in NC and CC modes, 
with a much lower energy threshold and better L/E resolution than 
SuperKamiokande. Actually, a 600 ton ICARUS liquid Argon detector will already
be installed in Gran Sasso by the year 2001, and despite the smaller mass
it will be able to cover the SuperK allowed region, due to the smaller energy
threshold.
\begin{figure}
\begin{center}
\epsfig{file=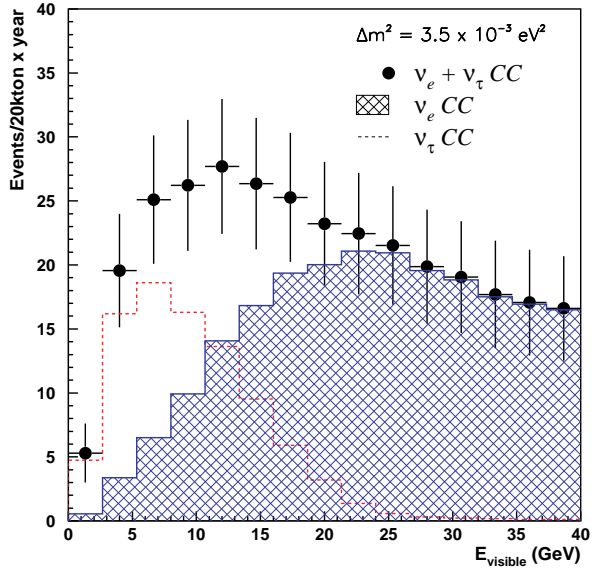,width=8cm,bb=45 150 450 540,clip}
\end{center}
\caption{Visible energy for $\nu_\tau$ events (red histogram) and $\nu_e$
background (dashed histogram). The sum of the two, with realistic statistical 
errors, is shown in crosses}
\label{fig:evis}
\end{figure}

\begin{figure}[tb]
 \begin{minipage}{.45\linewidth}
  \begin{center}
   \epsfig{file=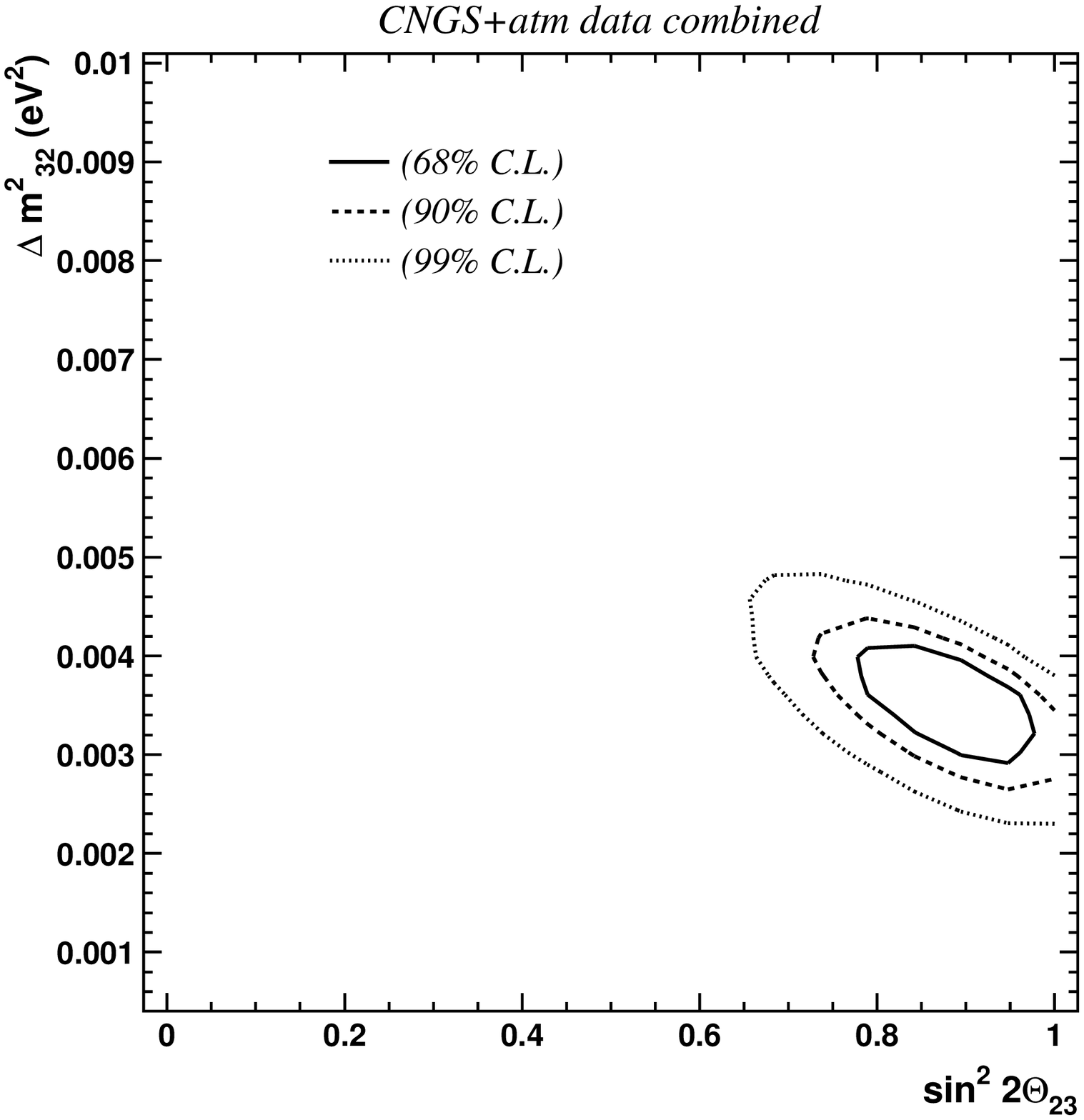,width=8cm}
  \end{center}
\caption{Precision on the measurement around the central values of 
$\Delta m^2_{23}=3.5\times 10^{-3} eV^2$ and $\sin^2\theta_{23}=0.9$.}
\label{fig:icapara1}
\end{minipage}
 \begin{minipage}{.45\linewidth}
  \begin{center}
   \epsfig{file=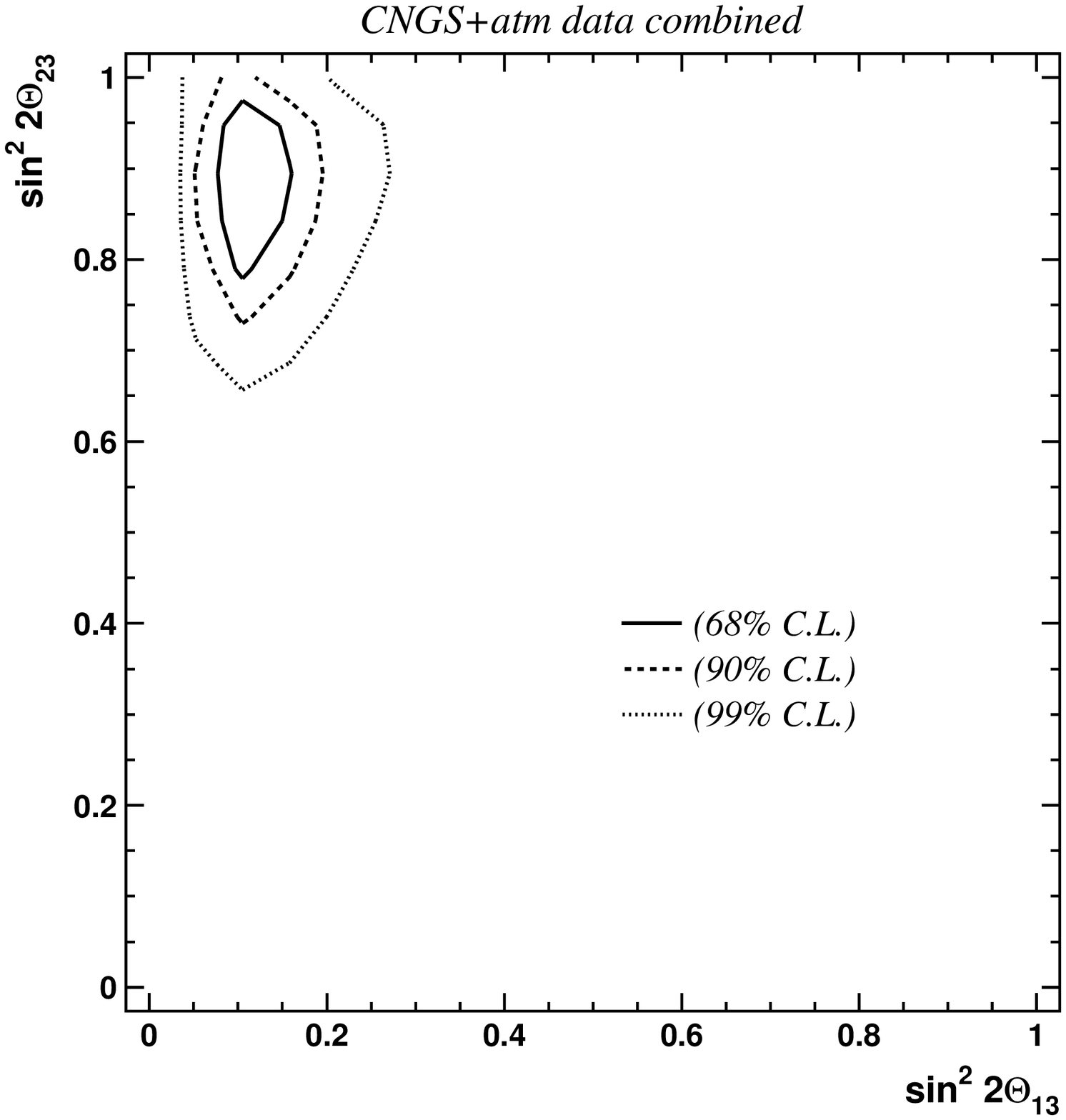,width=8cm}
  \end{center}
\caption{Precision on the measurement around the central values of 
$\sin^2 2\theta_{13}=0.10$ and $\sin^2\theta_{23}=0.9$.}
\label{fig:icapara2}
\end{minipage}
\end{figure}

Another main physics topic is of course the study of neutrinos from the CNGS
beam. This detector is able to perform $\nu_\mu\to\nu_e$ and 
$\nu_\mu\to\nu_tau$ searches, exploiting the different kinematics
of electrons from $\tau\to e$ decays and from electron neutrinos in the beam.
As we can see in figure \ref{fig:evis}, already at the level of total visible
energy the presence of $\tau$ neutrinos can be detected, and after kinematical
cuts 37 $\nu_\tau$ events are expected to be seen against a background of
4.4 for $\Delta m^2_{23}=3.5\times 10^{-3} eV^2$ after 4 years of data taking.
Combining results from atmospheric neutrinos with those from the CNGS
beam, it is possible to derive measurements of the oscillation parameters of
the order of 10\% for a reasonable choice of the parameters (Fig.
\ref{fig:icapara1} and \ref{fig:icapara2}). 
\subsection{Large atmospheric neutrino detectors}
\begin{figure}[tbh]
  \begin{center}
   \epsfig{file=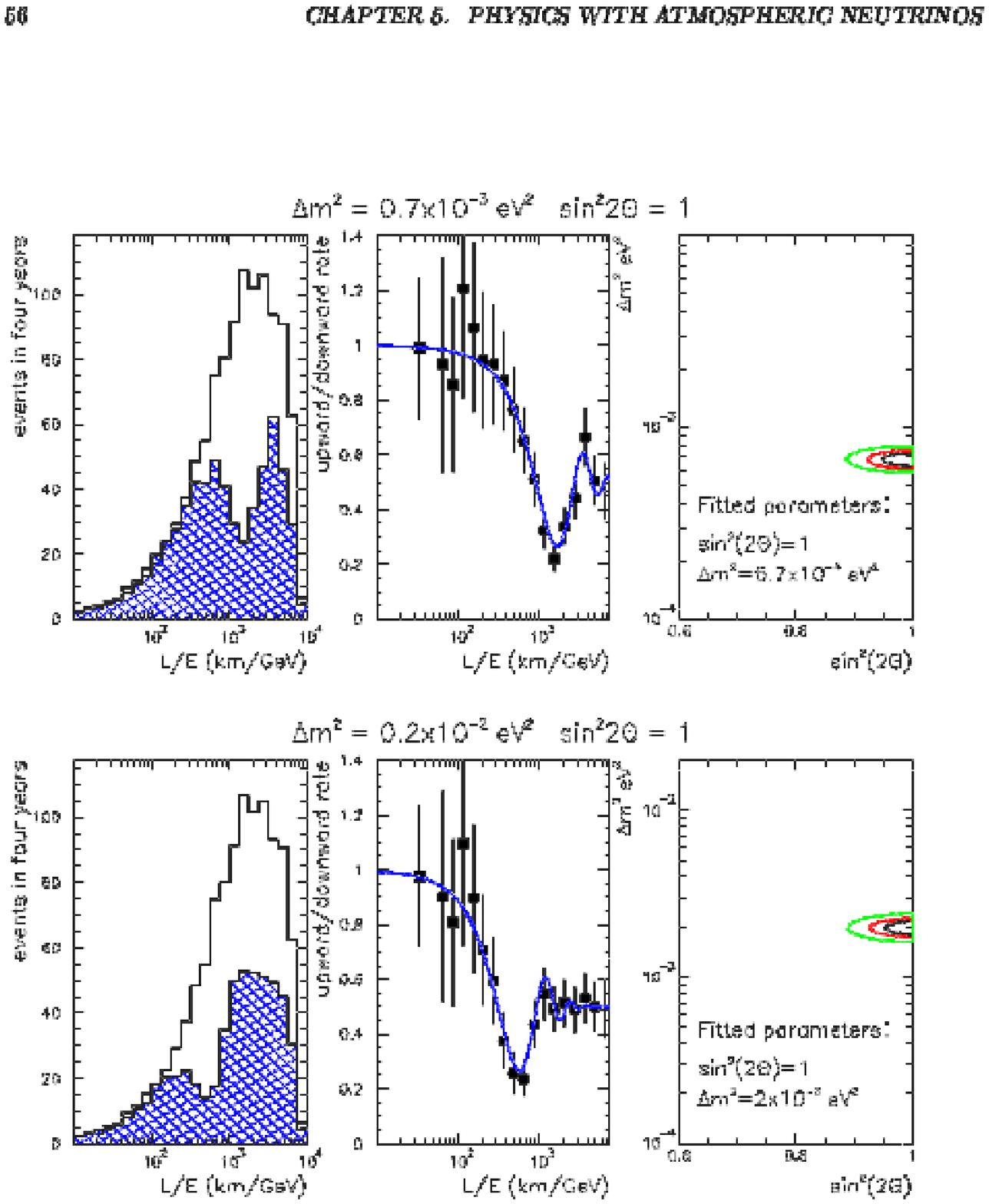,width=11cm,height=8cm,bb=60 80 520 620,clip}
  \end{center}
\caption{For two different values of $\Delta m^2_{23}$, from left to right:
the energy spectrum without and with oscillations; the ratio of the two;
the precision for measurement of the oscillation parameters.}
\label{fig:dipmono}
\end{figure}

A post-SuperKamiokande atmospheric neutrino detector should have a better L/E
resolution, to allow the observation of the oscillation dip in
atmospheric events. This is the idea beneath the Monolith proposal
for Gran Sasso \cite{monol}, a large (34 kton) coarse calorimeter (120 8 cm
thick iron plates alternated with 2 cm of gas spark chambers) with a 1T
magnetic field. Given the coarseness of the apparatus, the detection of the
hadronic part in charged current interactions is very difficult, and
a good determination of the initial neutrino direction can be obtained only
at high energy, where the correlation between neutrino and muon direction is
quite strong. For this reason a 1.5 GeV cut on the muon energy 
is applied, that limits the statistics to about 7 events/kton/year.
The large mass of the detector allows however to see the oscillation dip
down to values of $\Delta m^2_{23}\approx 10^{-4} eV^2$ (figure 
\ref{fig:dipmono}).\par
Another possible solution for a next-generation atmospheric neutrino detector
would be a follow-up of Super-Kamiokande, i.e. a giant (1 Mton) water Cerenkov.
The interest in such a detector is mainly driven by the search for
proton decay in the classic channel $p\to e^+ pi^0$. The study of atmospheric
neutrinos would clearly be another main reason for building such a detector;
however, already for the present SuperK data the systematic error has the same
magnitude as the statistical error; moreover, price reasons would force a 
smaller photo-multiplier coverage, that could result in poorer particle
identification capabilities, and still to be solved are the problems connected
with the large excavation time needed and the price of such an apparatus.

\begin{figure}[tbh]
  \begin{center}
   \epsfig{file=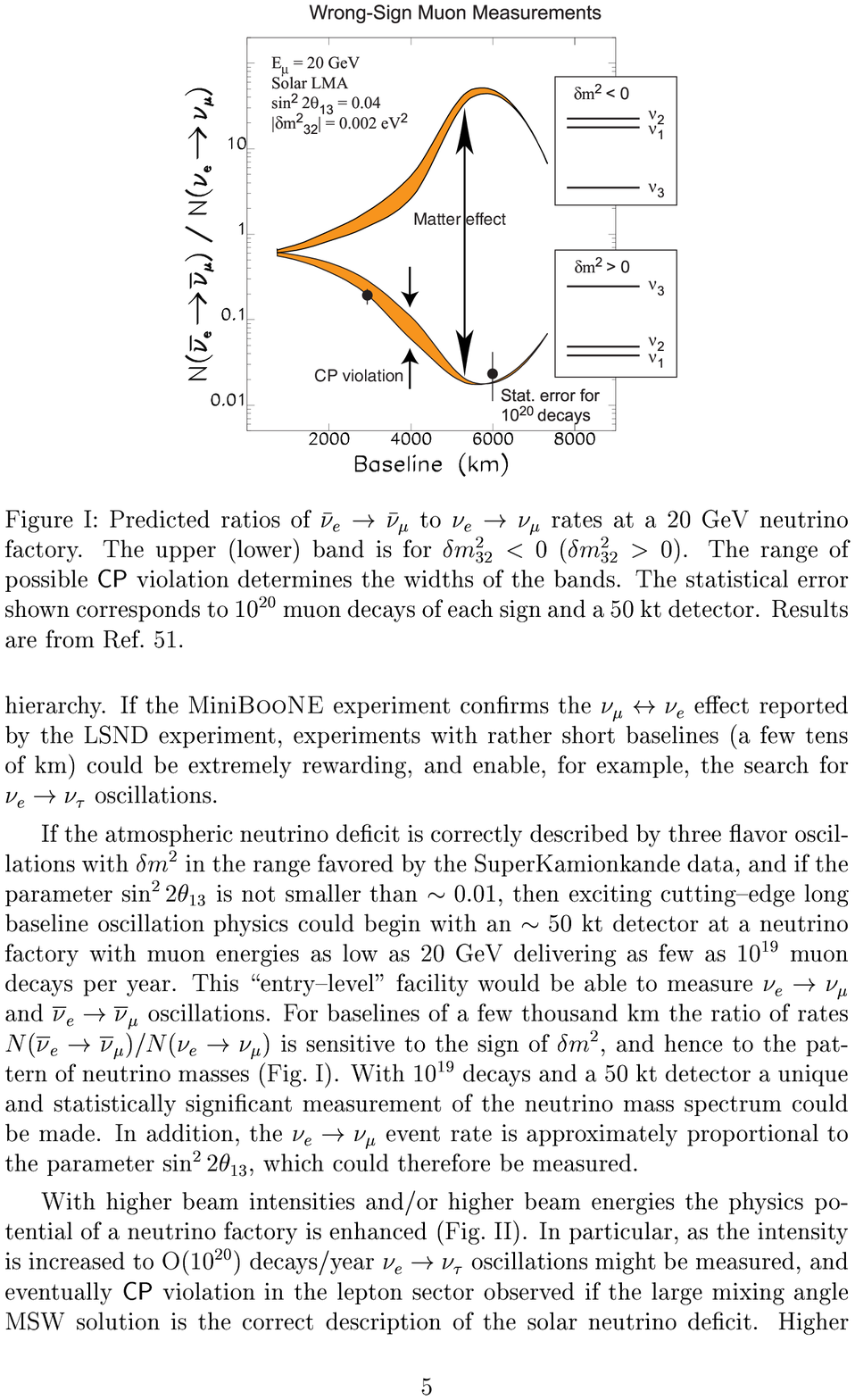,width=8cm,bb=175 457 440 720,clip}
  \end{center}
\caption{The ratio in the number of wrong-sign muons between runs with positive
and negative muons in the ring. The two curves represent the two signs of
$\Delta m^2_{23}$, and the tickness of the line represent the variation due
to different values of the CP-violationg parameter $\delta_{13}$.}
\label{fig:cpmsw}
\end{figure}

\section{Looking at the future}
In 10 years from now, if no particular surprise emerges, the situation of
neutrino oscillation could be as follows:\begin{itemize}
\item oscillations should be confirmed with an artificial beam by K2K, and the
oscillation parameters measured with better accuracy by MINOS
\item $\tau$ neutrinos from a $\nu_\mu$ beam should be observed by the two
CNGS experiments
\item $\nu_\mu\to\nu_e$ oscillation could be observed by ICANOE, or a limit
on $\sin^2 2\theta_{13}<10^{-3}$ could be derived
\item $\sin^2\theta_{23}$ and $\Delta m^2_{23}$ measured with 10\% precision
\end{itemize}
Goals for a next generation experiment would be:\begin{itemize}
\item measurement of the parameters of the leading oscillation with O(1\%)
precision
\item improvement of sensitivity on $\theta_{13}$, i.e. on $\nu_mu\to\nu_e$ 
oscillations
\item observation of matter effects on earth
\item discovery of CP violation in the leptonic system.
\end{itemize}
To reach these ambitious goals, new tools are needed. In particular, many
studies are going on to assess the physics potentials of a Neutrino Factory,
\cite{nufact} i.e. a machine where neutrino beams would be created from the 
decay of muons in a storage ring. The main advantage of this approach as 
opposed to ``traditional'' neutrino beams from proton decays are the fact that
neutrino beams from muons would have only two flavors of different helicity
and well-known spectra, while beams from pion decay have all flavors (even a
small $\nu_\tau$ component), and their spectra are affected by uncertainties
in hadronic production. In a neutrino factory, both muon charges would be
possible, leading to the decays $\mu^-\to e^-\bar{\nu}_e\nu_\mu$,
$\mu^+\to e^+\nu_e\bar{\nu}_\mu$, the muon energy could be tuned to have most
of the events in an interesting region, and the beam could be polarized to
enhance signal and control systematics. Since this machine is also
envisaged as a first step towards a muon collider, high intensities are
planned, of the order of $10^{20}$ muon decay per year. Such a high flux would
assure a reasonable statistics also for very long baseline experiments, i.e.
where the distance between neutrino production and detection is several 
thousands of kilometers. Given the atmospheric parameters, this would allow
observing the dip of $\nu_\mu\to\nu_\tau$ oscillations at an energy of about
15 GeV, and perform a precision measurement of its oscillation parameters.
A much improved precision can be reached in the search for $\nu_e\to\nu_\mu$
oscillations, looking for wrong-sign muons in the final state, i.e. muons
with opposite charge with respect to those circulating in the ring. These
events can only be originated by oscillations (with a very small small 
background from meson decays), in particular from the electron neutrino
component of the beam. In this case, the unique possibility offered by the 
neutrino factory of running with either positive or negative muons in the ring
offers the opportunity to study matter effects on earth \cite{msw},
given the different interference term of electron neutrinos or 
antineutrinos interacting with the electrons inside the earth. This difference
allows the determination of the sign of $\Delta m^2_{23}$, while all 
oscillations in vacuum or at shorter baselines are only sensitive to its
absolute value. The spectrum of wrong-sign muons is also distorted by the
presence of a complex term in the lepton mixing matrix, i.e. by the presence
of CP violation in the leptonic sector. Figure \ref{fig:cpmsw} shows the
asymmetry in the number of wrong-sign muon events due to matter effects and
CP violation as a function of the chosen baseline.
\section{Conclusions}
Even after the latest results in experimental neutrino physics, much has still
to be done to have a clear view of the mixing in the leptonic sector.
The first aspect to be clarified is the validity of the LSND result, that
still awaits an independent confirmation, and would need the introduction of
sterile neutrinos for its interpretation. The MiniBOONE experiment will
explore this region, expecting large signals in case of validity of the LSND 
result, even if also backgrounds will be large.\par
The atmospheric region will be explored by long baseline neutrino beams in
Japan, United States and Europe. They will confirm the disappearance of muon
neutrinos (K2K), measure the oscillation parameters (MINOS), and confirm
that the oscillation is mainly due to $\nu_\mu\to\nu_\tau$ transitions
(OPERA/ICANOE). The ICANOE detector will also play a leading role to the search
for $\nu_\mu\to\nu_e$ transitions.\par
After these experiments, a next generation of machines will be needed, to 
improve the precision of the measurements, to have more sensitivity on 
$\theta_{13}$ or discover $\nu_e\to\nu_\mu$ transitions, to study matter 
effects on earth and to possibly discover CP violation in the leptonic sector.
For all these studies, the best solution is probably to build a Neutrino
Factory where neutrinos are produced from the decay of stored muons. The high
neutrino fluxes considered are essential for precision measurements, as well
as for permitting very-long baseline experiments, that allow a detailed study
of matter effects, and would open a window on CP violation in the leptonic
sector.

\section*{Acknowledgments}
I would like to thank the organizers of the IV Rencontres du Vietnam for 
having invited me to give this talk, and for having given me the opportunity 
of discovering this fascinating country.

\end{document}